\documentclass[10pt,english,aip,twocolumn,reprint]{revtex4-2}
\usepackage[T1]{fontenc}
\usepackage{color}
\usepackage{amsmath}
\usepackage{amssymb}
\usepackage{graphicx}

\makeatletter
\usepackage[colorlinks, linkcolor=blue, anchorcolor=red, citecolor=blue]{hyperref}
\usepackage{graphicx}
\usepackage{bm}

\usepackage{mathptmx}
\usepackage{etoolbox}

\makeatother

\usepackage{babel}
\begin{document}

\global\long\def\bp{\boldsymbol{p}}%
\global\long\def\bv{\boldsymbol{v}}%
\global\long\def\be{\boldsymbol{e}}%
\global\long\def\bk{\boldsymbol{k}}%
\global\long\def\bx{\boldsymbol{x}}%
\global\long\def\ee{\mathrm{e}}%
\global\long\def\bnabla{\boldsymbol{\nabla}}%
\global\long\def\ii{\mathrm{i}}%
\global\long\def\d{\mathrm{d}}%

\title{Exact Normal Modes of Quantum Plasmas}
\author{Tian-Xing Hu}
\affiliation{Key Laboratory for Laser Plasmas and School of Physics and Astronomy,
and Collaborative Innovation Center of IFSA, Shanghai Jiao Tong University,
Shanghai, 200240, China}
\author{Dong Wu}
\email{dwu.phys@sjtu.edu.cn}
\affiliation{Key Laboratory for Laser Plasmas and School of Physics and Astronomy,
and Collaborative Innovation Center of IFSA, Shanghai Jiao Tong University,
Shanghai, 200240, China}

\author{Z. M. Sheng}
\affiliation{Institute for Fusion Theory and Simulation, Department of Physics,
Zhejiang University, Hangzhou 310027, China.}

\author{J. Zhang}
\email{jzhang1@sjtu.edu.cn}
\affiliation{Key Laboratory for Laser Plasmas and School of Physics and Astronomy,
and Collaborative Innovation Center of IFSA, Shanghai Jiao Tong University,
Shanghai, 200240, China}
\affiliation{Institute of Physics, Chinese Academy of Sciences, Beijing 100190, China.}

\begin{abstract}

The normal modes, i.e., the eigen solutions to the dispersion relation equation, 
are the most fundamental properties of a plasma, which also of key importance to many nonlinear effects such as parametric and two-plasmon decay, and Raman scattering. The real part indicates the intrinsic 
oscillation frequency while the imaginary part the Landau damping rate.
In most of the literatures, the normal modes of quantum plasmas are obtained by means of small damping approximation (SDA),
which is invalid for high-$k$ modes.
In this paper, we solve the exact dispersion relations via the analytical continuation (AC) 
scheme, and, due to the multi-value nature of the Fermi-Dirac distribution,
reformation of the complex Riemann surface is required.
It is found that the change of the topological shape of the root locus in quantum plasmas is quite different from classical plasmas,
in which both real and imaginary frequencies of high-$k$ modes increase with $k$ in a steeper way than the typical linear behaviour as appears in classical plasmas.
As a result, the temporal evolution of a
high-$k$ perturbation in quantum plasmas is dominated by the ballistic
modes. 

\end{abstract}

\maketitle

\section{Introduction}

The study of quantum plasmas has important applications in the fields
of, e.g., astrophysics\citep{Uzdensky_2014}, nano-physics\citep{manfredi2019phase,li2013landau},
warm dense matter\citep{PhysRevLett.77.1496}, and inertial confinement
fusion\citep{zhang2020double,PhysRevE.102.033312,ning2022}, and thus
received widespread attention in recent years. However, some of its
fundamental properties have rarely been seriously discussed. For example,
the exact eigen solution of a degenerate quantum plasma. In this
paper, we solve the Landau damping rate by means of analytical continuation
(AC) method, and avoid the branch cut discontinuity by extending the
Riemann surface of the 1d Fermi-Dirac distribution function (1dFDDF).
The AC method is a crucial step to comprehend the damping normal modes of plasmas,
proposed by Landau a long time ago, which is widely used in the field of classical plasmas,
but rarely discussed for quantum plasmas.
The AC method is able to solve the exact normal modes with arbitrarily
high wave numbers, which are important when mirco-scale quantum kinetic
effects are involved. And, the exact linear dispersion
relation of real frequency helps us better understand nonlinear phenomena,
namely, the coupling between different linear modes, and the exact
imaginary frequency determines the life-time of a plasmon.

The theoretical basis of this paper is the so-called collision-less
quantum kinetic theory (QKT), which is, mathematically speaking, a
Wigner-Poisson system (WPS) of equations. The linearized WPS is equivalent
to the famous Random Phase Approximation (RPA), or the time-dependent
Hartree approximation, which is well-proved to be a very successful
model for quantum plasmas when the density is higher than the typical
solid density ($\sim10^{24}\text{cm}^{-3}$). Such a dense, degenerate
electron environment is ubiquitous in the universe. For example, the
core of the sun is partially degenerate, the core of a white dwarf
is extremely degenerate, and, the progenitors of most of energetic
astrophysics events such as X-/$\gamma$-ray burst and supernova explosion,
are assumed to be related to extreme environments\citep{piran2005physics,seward2010exploring}.
In laboratories, some of the inertial confinement fusion experiment
also produce degenerate quantum plasmas, e.g., the double-cone ignition\citep{zhang2020double}
(DCI) scheme. Hence, this work may has a wide range of applications.

This paper is organized as follow.
In Sec. \ref{TM}, we introduced the complex structure of the Fermi-Dirac distribution function,
then briefly reviewed the collisionless QKT and the corresponding linear response theory,
which is the theoretical basis of this paper.
In Sec. \ref{QLW}, we discussed the complex structure of the dielectric function of quantum plasmas.
And thus the exact linear dispersion relation of quantum Langmuir wave is solved
by means of the aforementioned methods. Numerical simulations are also presented to verify the results.
Further discussion and main conclusion are presented in Sec. \ref{DC}.

\section{Theories and Methods}\label{TM}

We define a quantum parameter
\begin{equation}
\hat{\hbar}\equiv\frac{\hbar\omega_{\mathrm{p}}}{2E_{\mathrm{F}}},
\end{equation}
sometime is refereed to as the normalized Planck's constant. Here,
$\omega_{\mathrm{p}}=\sqrt{4\pi e^{2}n/m_{e}}$ is the plasma frequency,
and the Fermi energy $E_{\mathrm{F}}=\hbar^{2}\left(3\pi^{2}n\right)/2m_{e}$.
It seems strange that $\hat{\hbar}$ is proportional to $n^{-1/2}$, since we expect that
quantum effects should be stronger for higher density. However, quantum
effects are also stronger for lower temperature, while $\hat{\hbar}$
solely dependent on $n$. Hence, to measure the
importance of quantum wave effects, a more appropriate choice would be 
\begin{equation}
\tilde{\hbar}\equiv\frac{\hbar\omega_{\mathrm{p}}}{2k_{\mathrm{B}}T},
\end{equation}
and once the degeneracy $\Theta=k_{\mathrm{B}}T/E_{\mathrm{F}}$
(the inverse of which measures the importance degeneracy) is fixed,
decreasing $\hat{\hbar}$ means increasing $\tilde{\hbar}$.

In this paper, we adopt the natural unit system, where $\hbar=m_{e}=e=k_{\mathrm{B}}=1$.
And the frequency, number density, velocity, length, and energy, are
normalized to $\omega_{\mathrm{p}}$, $n_{e}$ (electron density),
$v_{\mathrm{F}}$ (Fermi velocity), $\lambda_{\mathrm{F}}=v_{\mathrm{F}}/\omega_{\mathrm{p}}$,
and $E_{\mathrm{F}}$ respectively.

\subsection{Analytical structure of the 1dFDDF}

The equilibrium state of electron obeys the 3d Fermi-Dirac distribution
(3dFDDF) 
\begin{equation}
\begin{aligned}f_{\mathrm{3d}}(\boldsymbol{v}) & =\frac{3}{4\pi}\frac{1}{\mathrm{e}^{\left(v^{2}-\mu\right)/\Theta}+1}\\
 & =-\frac{3}{4\pi}\mathrm{Li}_{0}\left[-\mathrm{e}^{\left(\mu-v^{2}\right)/\Theta}\right],
\end{aligned}
\end{equation}
where $\mathrm{Li}_{n}$ is the $n$-th order polylogarithm, and the
value of the chemical potential $\mu=\mu\left(\Theta\right)$ is chosen
such that $\int f_{\mathrm{3d}}(v)\d^{3}v=1$.

\begin{figure}
\begin{centering}
\includegraphics[width=0.95\columnwidth]{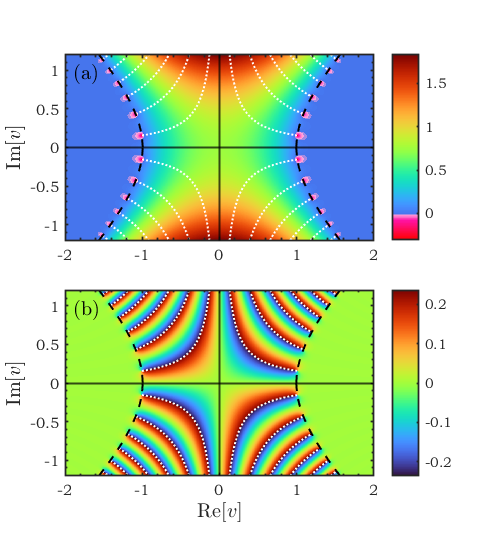}
\par\end{centering}
\caption{(a) Real (b) Imaginary part of the 1dFDDF ($\Theta=0.1$). The dash-lines
are the hyperbolae of Eq. (\ref{hyperbola}), and the white dotted-lines
are the branch cuts of Eq. (\ref{branch_cut}).}
\label{1dfddf}

\end{figure}

\begin{figure*}[tp]
\begin{centering}
\includegraphics[width=0.95\linewidth]{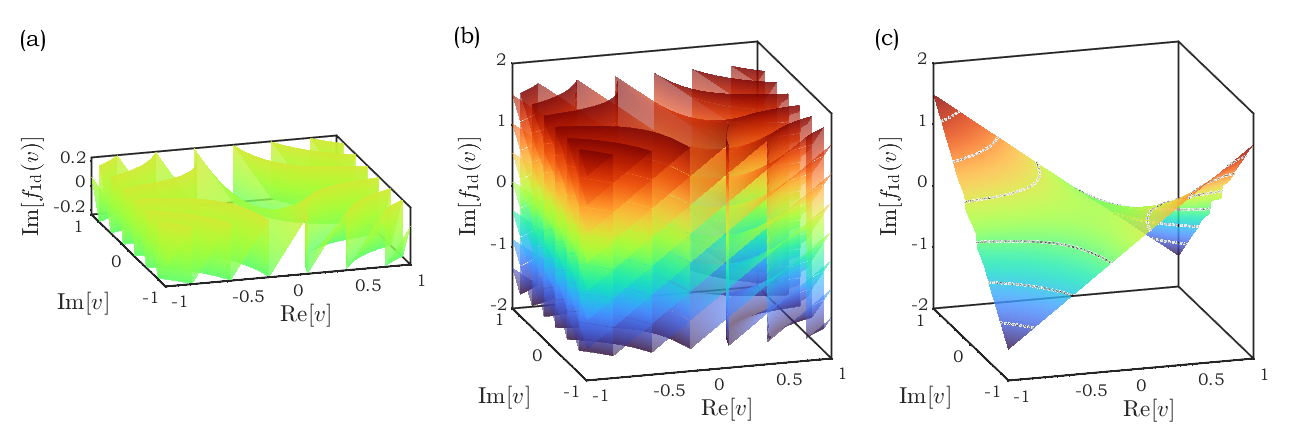}
\par\end{centering}
\caption{(a) A single leaf (b) Seven leaves (c) Extended single leaf of the
1dFDDF ($\Theta=0.1$) Riemann surface.}
\label{riemann3d}

\end{figure*}

As a fundamental study, we only consider the interaction of electrons
to a plane wave field, hence we can integrate over the dimensions
perpendicular to the wave, namely, we only care the about the 1d Fermi-Dirac
distribution (1dFDDF) $f_{\mathrm{1d}}(v_{\parallel})=\iint\d\bv_{\perp}f_{\mathrm{3d}}(v_{\parallel},\bv_{\perp})$.
It is easy to prove that,
\begin{equation}
\begin{aligned}f_{\mathrm{1d}}(v) & =-\frac{3}{4}\Theta\mathrm{Li}_{\frac{1}{2}}\left[-\mathrm{e}^{\left(\mu-v^{2}\right)/\Theta}\right]\\
 & =\frac{3}{4}\Theta\ln\left[\mathrm{e}^{\left(\mu-v^{2}\right)/\Theta}+1\right].
\end{aligned}
\label{eq:1dfddf}
\end{equation}
Let $v$ be a complex variant $v=v_{\mathrm{r}}+\mathrm{i}v_{\mathrm{i}}$,
then the 1dFDDF is a multi-value function, the branch cuts of which
are
\begin{equation}
v_{\mathrm{r}}v_{\mathrm{i}}=\pm\frac{\pi}{2}\ell\Theta,\quad\ell\in2\mathbb{Z},\label{branch_cut}
\end{equation}
and the branch points are located at the hyperbola:
\begin{equation}
v_{\mathrm{r}}^{2}-v_{\mathrm{i}}^{2}=\mu.\label{hyperbola}
\end{equation}
 The structure of the 1dFDDF is also thoroughly discussed in Ref.
\onlinecite{vladimirov2011description}. The density of the branch
points on the hyperbola increases with decreasing $\Theta$. In Fig.
\ref{1dfddf}, we plotted the 1dFDDF in complex-$v$ plane, one can
see that its imaginary part is discontinued at the branch cuts. These
discontinuities are equal to the height of a Riemann leaf, which is
$3\pi\Theta/2$ here, and they exist because we considered only a
single leaf the Riemann surface, see Fig. \ref{riemann3d} (a). However,
the Riemann surface of 1dFDDF has an infinite leaves because of the
logarithm in Eq. (\ref{eq:1dfddf}), see Fig. \ref{riemann3d} (b).
Discontinuity may cause unphysical effects, as we shall see in the
next subsection. Hence, we extend the area enclosed by the first branch
cut and the hyperbola to the whole Riemann surface, as is shown in
Fig. \ref{riemann3d} (c), we then obtain a smooth surface between
the two branches of the hyperbola. Noticing that, after this operation,
the discontinuities do not vanish, but are moved from the branch cuts
to the hyperbola.

\subsection{The Linearized WPS}

The electrons in a collisionless quantum plasma obey the Wigner equation \citep{kadanoff2018quantum}
\begin{equation}
\partial_{t}f+\bv\cdot\partial_{\bx}f+\ii\vartheta_{\ii\hat{\hbar}\partial_{\bp}}[\phi\left(\bx\right)]f=0.\label{wigner}
\end{equation}
Here $f=f(\bx,\bv,t)$ is the Wigner quasi-distribution function,
which is the quantum counter part of the classical distribution function,
and the pseudo-differential operator $\vartheta$ is defined by
\begin{equation}
\vartheta_{\boldsymbol{y}}[O\left(\bx\right)]\equiv O\left(\bx+\frac{\boldsymbol{y}}{2}\right)-O\left(\bx-\frac{\boldsymbol{y}}{2}\right).
\end{equation}
Noticing that when $y\ll x$, $\vartheta_{\boldsymbol{y}}[O\left(\bx\right)]\simeq\boldsymbol{y}\cdot\bnabla O$,
then Eq. (\ref{wigner}) reduces to the Vlasov equation, thus the
Wigner equation is also called the quantum Vlasov equation.

And we need the Poisson equation
\begin{equation}
-\bnabla^{2}\phi=n_{\mathrm{b}}-e\int f\mathrm{d}\bv,\label{Poisson}
\end{equation}
where $n_{\mathrm{b}}$ stands for the background ion density, to
close the system. Let $f=f_{0}+\delta f$, $\phi=\phi_{0}+\delta\phi$,
and consider only the direction parallel to the wave vector, the linear
evolution of the perturbed field can be formally written as \citep{chen1987waves}
\begin{equation}
\delta\phi\left(t,k\right)=\frac{\ii}{k^{2}}\iint\d v_{\parallel}\frac{\d\omega}{2\pi}\frac{\delta f_{k0}\left(v_{\parallel}\right)\ee^{-\ii\omega t}}{\left(\omega-kv_{\parallel}\right)\epsilon\left(\omega,k\right)},\label{deltaphi}
\end{equation}
where $\delta f_{k0}\left(v\right)$ is the initial perturbation,
and the dielectric function (DF) is defined as
\begin{equation}
\epsilon\left(\omega,k\right)=1+\frac{1}{k^{2}}\mathcal{W}\left(\hat{\hbar}\omega,\hat{\hbar}k\right),\label{df}
\end{equation}
where

\begin{equation}
\mathcal{W}\left(\omega,k\right)=\int\d v\frac{\vartheta_{k}\left[f\left(v\right)\right]}{\omega-kv}\label{Lindhard}
\end{equation}
is the Lindhard response function \citep{lindhard1954properties}
(LRF). As $\Theta\rightarrow\infty$, $f(v)$ tends to Maxwellian,
namely, in the classical limit, it reduces to
\begin{equation}
\mathcal{W}\left(\omega,k\right)=\frac{2}{\Theta}\frac{1}{\sqrt{2\pi}}\int\d v\frac{v\ee^{-\frac{v^{2}}{2}}}{v-\omega/k_{\Theta}}=-\frac{2}{\Theta}Z'\left(\frac{\omega}{k_{\Theta}}\right),
\end{equation}
where $k_{\Theta}=k\sqrt{\Theta/2}$, and $Z$ is the famous plasma
dispersion function. It is evident that in quantum plasma, the linear
response is dependent on both $\omega$ and $k$, but in classical
plasma, it depends only on the ratio of $\omega$ to $k$. The response
function then does not depend on $\hat{\hbar}$, which means that
when $\Theta$ is large enough, the system naturally returns to classical,
no matter the value of $\hat{\hbar}$.

The roots of the eigen-equation
\begin{equation}
\epsilon\left(\omega,k\right)=0\label{eigen-eq}
\end{equation}
yield the dispersion relation of the normal modes. Here, $\omega=\omega_{\mathrm{r}}+\ii\omega_{\mathrm{i}}$
is a complex number. To calculate the DF (\ref{df}) exactly with
negative value of $\omega_{\mathrm{i}}$ and solve Eq. (\ref{eigen-eq}),
analytical continuation (AC) is needed for $\omega_{\mathrm{r}}=0$
is a branch cut, namely, when $\omega_{\mathrm{i}}<0$, the response
function (\ref{Lindhard}) should be modified to
\begin{equation}
\mathcal{W}\left(\omega,k\right)=\int\d v\frac{\vartheta_{k}\left[f\left(v\right)\right]}{\omega-kv}-2\pi\ii\vartheta_{k}\left[f\left(\frac{\omega}{k}\right)\right],\label{Lindhard-ac}
\end{equation}
where the $2\pi\ii$ term stems from the residue of the integrand
of Eq. (\ref{Lindhard}). In the classical limit, it is 
\begin{equation}
\frac{\Theta}{2}\mathcal{W}\left(\eta\right)=1-\sqrt{2}\eta F\left(\frac{\eta}{\sqrt{2}}\right)+\ii\sqrt{\frac{\pi}{2}}\eta\ee^{-\frac{\eta^{2}}{2}},\label{class-W}
\end{equation}
where $\eta=\omega/k_{\Theta}$, and 
\begin{equation}
F(x)=\ee^{-x^{2}}\int_{0}^{x}\ee^{t^{2}}\d t
\end{equation}
is the Dawson integral.

\section{Quantum Langmuir Wave} \label{QLW}

\subsection{Solving the Normal Modes}

The Landau damping rate is the negative imaginary part of the eigen-frequency.
To calculation the DF with negative imaginary frequency, analytical
continuation is needed. Noticing that for extremely degenerate plasmas,
analytical continuation is not needed when $k<k_{\mathrm{F}}=v_{\mathrm{F}}=\omega_{\mathrm{p}}\lambda_{\mathrm{F}}^{-1}$,
since the dielectric function has an analytic solution when $\Theta\rightarrow0$:
\begin{equation}
\begin{aligned}\mathrm{Re}[\epsilon]=1+ & \frac{3}{2k^{2}}\left[1+\frac{1}{2\hat{\hbar}k}\left(1-b_{-}^{2}\right)\ln\left|\frac{1+b_{-}}{1-b_{-}}\right|\right.\\
 & \left.-\frac{1}{2\hat{\hbar}k}\left(1-b_{+}^{2}\right)\ln\left|\frac{1+b_{+}}{1-b_{+}}\right|\right],
\end{aligned}
\end{equation}
\begin{equation}
\mathrm{Im}[\epsilon]=\frac{3\pi^{2}}{4\hat{\hbar}k^{3}}\ln\frac{1+\exp\left[\left(b_{+}^{2}-\mu\right)/\Theta\right]}{1+\exp\left[\left(b_{-}^{2}-\mu\right)/\Theta\right]},
\end{equation}

where
\begin{equation}
b_{\pm}=\frac{\omega}{k}\pm\frac{\hat{\hbar}k}{2}.
\end{equation}
Thus one can see than only when $\left|b_{\pm}\right|>1$, namely,
\begin{equation}
-k\pm\frac{\hat{\hbar}k^{2}}{2}<\omega<k\pm\frac{\hat{\hbar}k^{2}}{2},\label{damping area}
\end{equation}
the DF has finite imaginary part. The area enclosed by (\ref{damping area})
is referred as to the electron-hole excitation continuum (EHEC), in
which a plasmon delay into an electron and a hole. The above formula
is obtained based on the small damping approximation (SDA):
\begin{equation}
\lim_{y\rightarrow0}\frac{1}{x\pm\ii y}=\mathcal{P}\frac{1}{x}\mp\ii\pi\delta(x),\label{Pleji}
\end{equation}
which is incorrect when $\omega_{\mathrm{i}}$ is finite. In the following,
we directly solve the Eq. (\ref{Lindhard-ac}) without any approximation
scheme.

\begin{figure}
\begin{centering}
\includegraphics[width=0.95\linewidth]{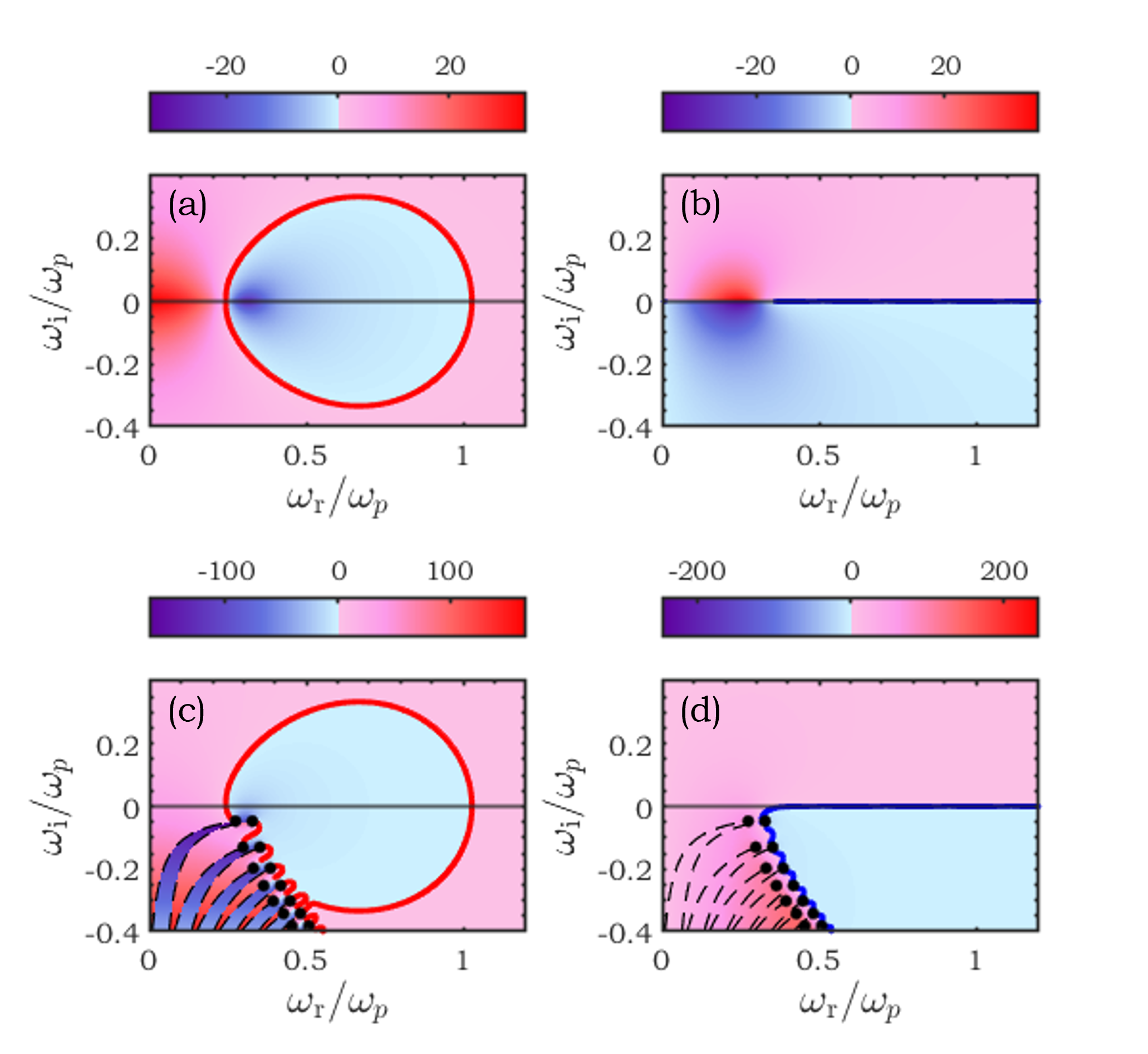}
\par\end{centering}
\caption{(a) (b): Real and imaginary part of the dispersion function without
analytical continuation, and (c) (d): with analytical continuation.
The dashed-lines are mappings of the branch cuts. The red lines stand
for $\mathrm{Re}\left[\epsilon\left(\omega,k\right)\right]=0$, and
blue lines for $\mathrm{Im}\left[\epsilon\left(\omega,k\right)\right]=0$.}
\label{continue}
\end{figure}

\begin{figure*}
\begin{centering}
\includegraphics[width=1\linewidth]{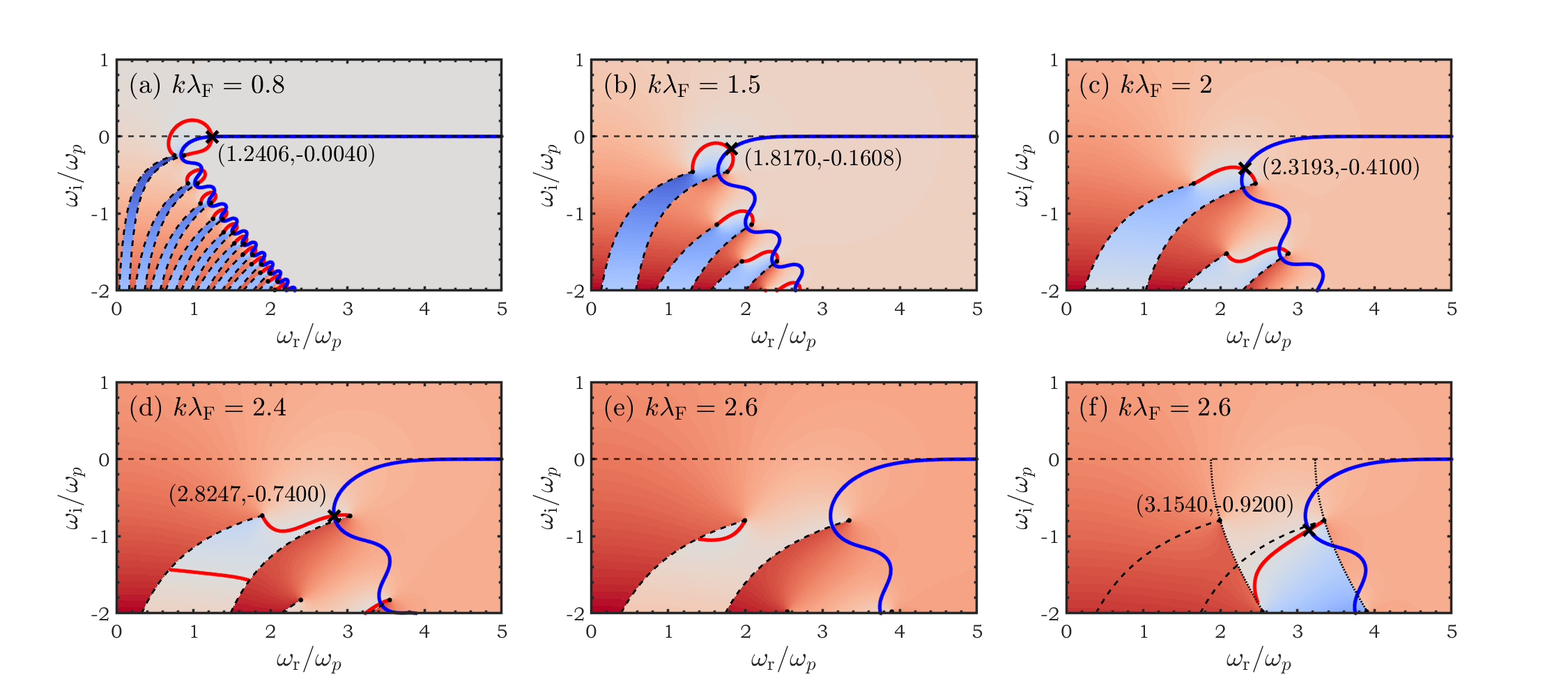}
\par\end{centering}
\caption{Complex frequency plane of dispersion function, with $\Theta=0.2,\hat{\hbar}=0.2$.}
\label{QLAN-1}
\end{figure*}

\begin{figure*}
\begin{centering}
\includegraphics[width=1\linewidth]{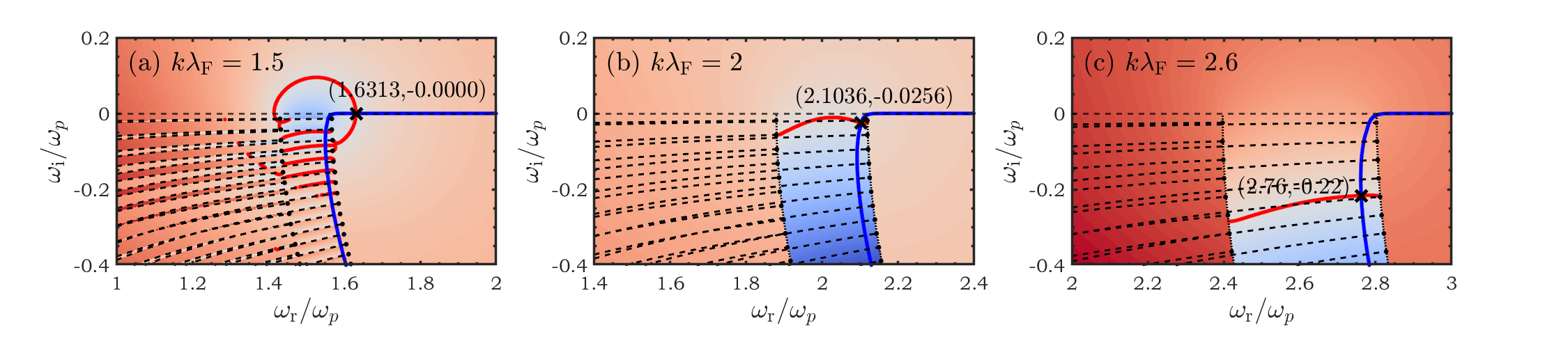}
\par\end{centering}
\caption{Complex frequency plane of dispersion function, with $\Theta=0.006,\hat{\hbar}=0.06$.}
\label{QLAN-2}
\end{figure*}

In Fig. \ref{continue}, the numerical results of dispersion function
with or without AC are presented ($\Theta=0.1,\hat{\hbar}=0.6$),
where the wave number $k=0.3$. The upper
two panels (a) and (b) are the real and imaginary parts of the DF
without AC, while the lower two are with. It is shown that the discontinuity
of the 1DFDDF is explicitly mapped from the complex $v$ plane to
the complex $\omega$ plane by the $\vartheta$ operator, since in
the calculation, we used the non-extended 1DFDDF, i. e., the single
leaf Riemann surface of Fig. \ref{riemann3d} (a). The series of pairs
of black dots in the lower half plane are located at two hyperbolas,
whose vertices are $\mu k\pm\hat{\hbar}k^{2}/2$ respectively.

In the rest of this paper, the imaginary part of the DF would not
be presented, but we still keep the $\mathrm{Im}\left[\epsilon\left(\omega,k\right)\right]=0$
lines in the real DF diagrams. And the colorbar will also be neglected
since the absolute values of the DF are irrelevant in the context. 

Take an example of a degenerate plasma, say, with density $n=3\times10^{26}\text{cm}^{-3}$,
and temperature $T=300\,\text{eV}$. Then we have the quantum parameter
$\hat{\hbar}=0.2$ and $\Theta=0.2$. The real DF of such parameters
in complex frequency plane are plotted In Fig. \ref{QLAN-1}, where
the real root loci are indicated by red lines and the imaginary loci
by blue lines. Noticing that the real and imaginary loci have multiple
intersections, which stand for multiple solutions of the normal modes.
Generally, we only care about the least damping mode, i.e., the intersection
point with maximum $\omega_{\ii}$, which we indicated by black crosses
in Fig. \ref{QLAN-1} and labeled its value. In panels (a) to (e),
we used the single leaf 1DFDDF, the discontinuities occur at the branch
cuts. One can see that there are no intersections in panel (e), so
we have to use the extended 1DFDDF to move the discontinuities to
the hyperbolas, as is seen in panel (f), in which a damping mode is
obtained.

\begin{figure}
\begin{centering}
\includegraphics[width=1\columnwidth]{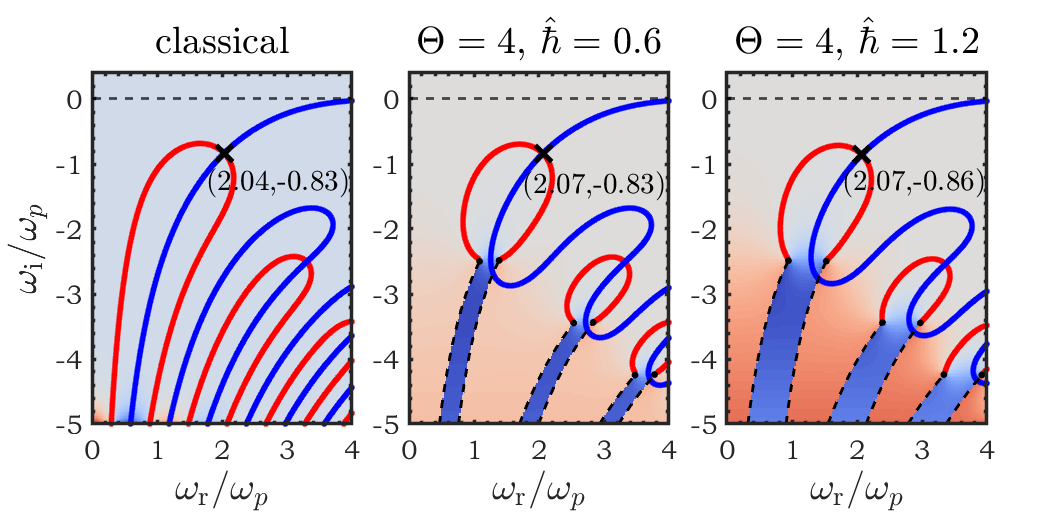}
\par\end{centering}
\caption{The DF complex frequency plane of classical and almost-classical plasmas.}
\label{q2c}
\end{figure}

For another example of extremely degenerate plasma, we choose density
$n=3\times10^{29}\text{cm}^{-3}$, and temperature $T=10^{3}\,\text{eV}$,
which could be the interior of a typical small white dwarf. Then we
have the $\hat{\hbar}=0.06$ and $\Theta=0.005$. The DF results are
presented in Fig. \ref{QLAN-2}. One finds that, at such small $\Theta$,
the distance among each branch cuts are extremely small. All the modes
with finite damping rate are located between the two hyperbolas (a
gap opened by quantum wave effect). Hence the extend 1DFDDF is indispensable
in this case. From Fig. \ref{QLAN-2}, one also finds that when $\omega_{\mathrm{i}}$
is finite,
\begin{equation}
\omega_{\mathrm{r}}\simeq\mu k+\frac{\hat{\hbar}k^{2}}{2},
\end{equation}
and since $\mu\rightarrow1$ when $\Theta\rightarrow0$, this is upper
bound of the EHEC.

Now we briefly discuss the topological features of the root loci.
Generally, the imaginary locus starting from $\infty+0\ii$, passing
through every pair of branch point, while the real locus starting
from one of the branch points and end with another. If the starting
and ending points are a pair, and the locus is above the line joining
the two branch points, we refer to this topological shape as ``classical'',
for it is topologically identical to the classical solution, see Fig.
\ref{q2c} (The classical thumb-like figure is very common in the
field of classical plasma instabilities, e.g., see Ref. \onlinecite{Feng2019}).
Otherwise it is a ``quantum'' shape. For example, the (a) \textasciitilde{}
(d) panels in Fig. \ref{QLAN-1} are classical, while (d), (e) and
(f) are quantum. And in extremely degenerate case like Fig. \ref{QLAN-2},
the classical shape does not exist at all. Furthermore, a quantum
shape means that the least damping mode is most likely located within
the two hyperbolas. 

In Fig. \ref{q2c}, the classical sub-figure is calculated via Eq.
(\ref{class-W}), and the other two are quantum results with $\Theta=4$,
which is almost Maxwellian, it is shown that the normal modes in these
there sub-figure are fairly close to one another. As $\Theta\rightarrow\infty$,
all the branch points going to $-\ii\infty$, and we know that the
value of $\hat{\hbar}$ (or $\tilde{\hbar}$) only effects the distance
between a pair of branch points. This diagrammatically demonstrated
that the larger $\Theta$ is, the less important the value of $\hat{\hbar}$
(or $\tilde{\hbar}$) is.

\subsection{Exact Solution of the Normal Modes}

\begin{figure*}
\begin{centering}
\includegraphics[width=1\linewidth]{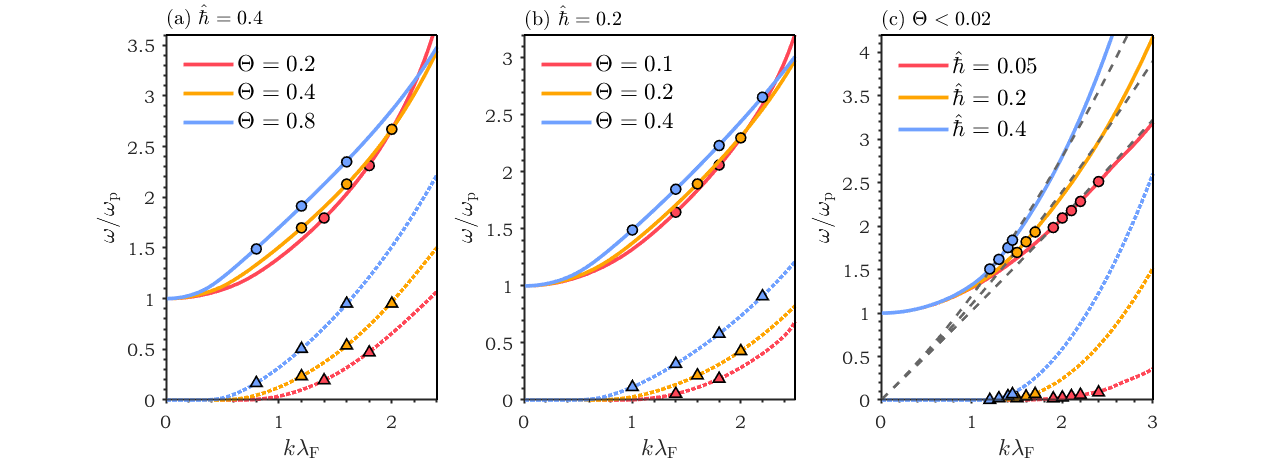}
\par\end{centering}
\caption{Dispersion relation calculated via AC method, where the solid lines
stand for real frequency and dashed-lines for imaginary, and the colored
markers are simulation results (square for real and triangle for imaginary
part). }
\label{eigen-solutions}
\end{figure*}

It is worth mentioning that, $\epsilon\left(\omega,k\right)=0$ has
multiple roots in a finite region of the complex $\omega$ plane,
as one can see from Fig. \ref{continue}, but we only care about the
lowest damping mode. Hence, to solve the full dispersion relation
of the QLW, we can starting from $k\simeq0^{+}$, and calculate a
small region centered at $\omega_{0}=1+0^{+}\ii$ to find the first
root $\omega_{1}$ and record the difference $\Delta\omega_{1}=\omega_{1}-\omega_{0}$.
Then increasing $k$ with a small value and calculating a next region
centered at $\omega_{1}+\Delta\omega_{1}$ to find the second root.
Repeating this procedure, then a continuous curve of the complex frequency
of the QLW as a function of $k$ is obtained. The full solutions of
normal modes $\omega\left(k\right)$ with different densities are
plotted in Fig. \ref{eigen-solutions}. In panel (a), where $\hat{\hbar}=0.4$
is corresponding to electron number density $n=4\times10^{24}\text{cm}^{-3}$,
and $\Theta=0.2$, $0.4$, and $0.8$, are corresponding to the temperature
$T=19$, $37$ and $74$eV. In panel (b), where $\hat{\hbar}=0.2$
is corresponding to electron number density $n=3\times10^{26}\text{cm}^{-3}$,
and $\Theta=0.1$, $0.2$, and $0.4$, are corresponding to the temperature
$T=170$, $340$ and $680$eV. In both the two cases, the eigen frequency
of lower temperature plasmas surpass the higher temperature when $k$
is large enough, which is the result of the quantum topological shape.
In panel (c), all the curves are calculated with $\Theta=0.02$, since
the shape of the curves hardly change when $\Theta$ further decreases,
we annotated they as $\Theta<0.02$, or, one can simply treat them
as zero-temperature results. The dashed-lines attached to each solid
line are the corresponding upper bounds of the EHEC. One can see that,
the solid lines are slightly lower than the dashed-lines only in a
very small region, which means that the EHEC predicted by Eq. (\ref{damping area})
is incorrect when $k$ is large. This is because the derivation of
Eq. (\ref{damping area}) is based on the SDA, which is incorrect
for finite damping rate.

\subsection{Numerical Benchmark of the Normal Modes}

The WPS, i. e., Eq. (\ref{wigner}) and (\ref{Poisson}) can be solved
numerically \citep{suh1991numerical,hu2022kinetic} as a initial value
problem. We set the initial perturbation as
\begin{equation}
f(x,v,0)=\left[1+A\cos\left(k_{0}x\right)\right]f_{\text{1d}}(v),
\end{equation}
and let it evolve, then measure the frequency and damping rate of
the perturbation. The factor $A$ is chosen to be very small to avoid
nonlinear effects. When the initial perturbation is set, the time-dependent
behavior of the perturbed field $\delta\phi$ consist of a ballistic
mode and an infinity number of normal mode, and this can be shown
by rewrite Eq. (\ref{deltaphi}) as
\begin{equation}
\begin{aligned}\delta\phi\left(t,k\right)= & \frac{1}{k^{2}}\int\d v\delta f_{k0}\left(v\right)\\
 & \times\left[\frac{\ee^{-\ii kvt}}{\epsilon\left(kv,k\right)}+\sum_{n}\frac{\ee^{-\ii\omega_{n}t}}{\left(\omega_{n}-kv\right)\partial_{\omega_{n}}\epsilon}\right],
\end{aligned}
\label{deltaphi2}
\end{equation}
where the first term in the bracket stands for the ballistic mode
while the second the normal modes.

Some results are presented also in Fig. \ref{eigen-solutions}, where
the colored circles stand for real frequency measured from simulation
result, and the triangles for damping rate. The simulation results
and the numerical results of normal modes match up perfectly. However,
the simulation results of high-$k$ modes are not shown, especially
in panel (c), where the damping rates of the simulation points with
the largest $k$ are one order lower than they real frequency. We
do not present the result of high $k$ because it is found that those
high-$k$ modes have uncertain frequency, or, do not damp exponentially,
it is impossible to measure the complex frequency nor the damping rate. The reason for
this non-exponential behavior is, as we have mentioned previously,
in degenerate plasmas, when $k$ is large enough, the topological
shape of the real locus become non-classical, which results in the
frequency and the damping rate of normal modes increase faster than
linear with $k$. Hence, beyond a critical value of $k$, the damping
rate of the normal mode will surpass the ballistic mode, the time-dependent
behavior then dominated by the ballistic mode. This phenomenon does
not occur in a classical plasma since its dispersion relation is always
linear as $k$ increases (when $k\lambda_{\mathrm{\mathrm{D}}}\gtrsim1$). 

\begin{figure}
\begin{centering}
\includegraphics[width=0.94\columnwidth]{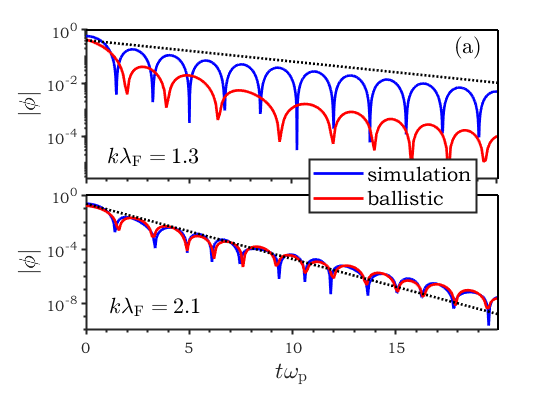}
\par\end{centering}
\begin{centering}
\includegraphics[width=0.94\columnwidth]{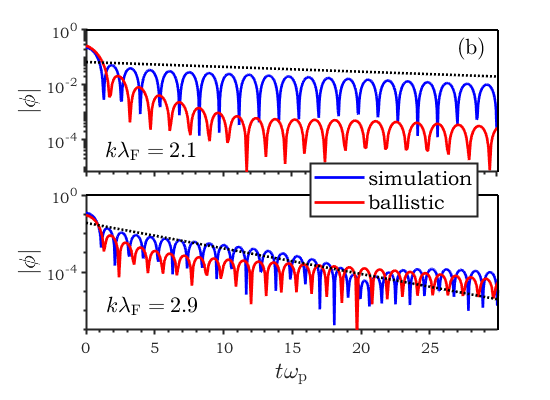}
\par\end{centering}
\caption{Temporal evolution of linear perturbations calculated by numerical
simulation (the blue lines), and the ballistic evolution (the red
lines): (a) $\Theta=0.2,\,\hat{\hbar}=0.6$, (b) $\Theta=0.02,\,\hat{\hbar}=0.05$.
The dotted-lines indicate the damping rate calculated by the eigen
equation (\ref{eigen-eq}).}
\label{simu_ballistic}
\end{figure}

The proof of the above statement is represented in Fig. \ref{simu_ballistic},
where the red lines are the pure ballistic evolution obtained by integrate
the first term in Eq. (\ref{deltaphi2}). In panel (a), where $\Theta=0.2$
and $\hat{\hbar}=0.6$, when $k=1.3$, the ballistic mode damps faster
than the normal mode, hence the simulation result shows a clear normal
mode with $\omega_{\mathrm{i}}=-0.185$ (the dotted line stands for
$\ee^{-0.185t}$). However, when $k=2.1$, the normal mode should
give $\omega_{\mathrm{i}}=-0.944$, while the simulation curve damps
slower than $\ee^{-0.944t}$ and is almost identical to the ballistic
mode, and it confirms our conclusion. Also noticing that, the simulation
curve does not have a clear real frequency. Similarly, in panel (b),
where $\Theta=0.02$ and $\hat{\hbar}=0.05$, a clear normal mode
with $\omega_{\mathrm{i}}=-0.0957\ii$ is measured for $k=2.1$, but
for $k=2.9$ the long time behavior is interfered by ballistic mode,
hence no clear normal mode can be seen.

\section{Discussion and Conclusion} \label{DC}

In this paper, we adopt the analytical continuation scheme to solve
the dispersion relation of degenerate plasma. Compared to SDA scheme,
the AC scheme can solve the normal modes of a dielectric system with
arbitrarily high $k$, which is crucial when small-wavelength quantum
kinetic effects are encountered. We find that in degenerate plasmas,
the normal mode frequency and its damping rate increase with $k$
steeper than linear, which is related to the ``quantum'' topological
shape of the root locus. As a result, the temporal evolution of a
high-$k$ perturbation in quantum plasmas is dominated by ballistic
mode. Especially, the exact solution of linear dispersion relation
is the basis of the quantitative analysis of nonlinear effects such
as parametric decay, two-plasmon decay, Raman scattering, and etc.,
in dense plasmas.

\section{Acknowledgements}
This work was supported by the Strategic Priority Research Program of Chinese Academy of Sciences (Grant No. XDA250050500), 
the National Natural Science Foundation of China (Grant No. 12075204), and Shanghai Municipal Science and Technology Key Project (No. 22JC1401500).
Dong Wu thanks the sponsorship from Yangyang Development Fund.

\bibliographystyle{unsrt}
\bibliography{ref}

\end{document}